\documentclass[11pt]{article}
\usepackage{ amsfonts, amssymb, graphicx, a4, epic, eepic, epsfig,
setspace, lscape}
\usepackage{color,soul,cancel}
\definecolor{Red}{cmyk}{0,1,1,0}
\definecolor{Blue}{cmyk}{1,1,0,0}
\definecolor{ForestGreen}{cmyk}{0.91,0,0.88,0.12}

\def\inst#1{$^{#1}$}

\newtheorem{theorem}{Theorem}[section]
\newtheorem{lemma}[theorem]{Lemma}
\newtheorem{proposition}[theorem]{Proposition}

\newtheorem{definition}[theorem]{Definition}
\newtheorem{corollary}[theorem]{Corollary}
\newtheorem{remark}[theorem]{Remark}

\newcommand{\cX}{{\cal X}}

\def \G {{\Gamma}}
\def \L {{\Lambda}}

\def \D {{\Delta}}
\def \r {{\rho}}

\def \m {{\mu}}

\def \s {{\sigma}}

\def \g {{\gamma}}
\def \t {{\tau}}

\def \d {{\delta}}
\def \p {{\pi}}

\def\<{{\langle}}
\def\>{{\rangle}}

\newcommand{\be}[1]{\begin{equation}\label{#1}}
\newcommand{\ee}{\end{equation}}

\newcommand{\bl}[1]{\begin{lemma}\label{#1}}
\newcommand{\el}{\end{lemma}}

\newcommand{\br}[1]{\begin{remark}\label{#1}}
\newcommand{\er}{\end{remark}}

\newcommand{\bt}[1]{\begin{theorem}\label{#1}}
\newcommand{\et}{\end{theorem}}

\newcommand{\bd}[1]{\begin{definition}\label{#1}}
\newcommand{\ed}{\end{definition}}

\newcommand{\bcl}[1]{\begin{claim}\label{#1}}
\newcommand{\ecl}{\end{claim}}

\newcommand{\bp}[1]{\begin{proposition}\label{#1}}
\newcommand{\ep}{\end{proposition}}

\newcommand{\bc}[1]{\begin{corollary}\label{#1}}
\newcommand{\ec}{\end{corollary}}

\newcommand{\bi}{\begin{itemize}}
\newcommand{\ei}{\end{itemize}}

\newcommand{\ben}{\begin{enumerate}}
\newcommand{\een}{\end{enumerate}}

\def \Z {{\mathbb Z}}

\begin {document}

%%%%%%%%%%%%%%%%%%%%%%%%%%%%%%%%%%%%%%%%%%%%%%%

%%%%%%%%%%%%%%%%%%%%%%%%%%%%%%%%%%%%%%%%%%%%%%%%%

\title{Probabilistic Cellular Automata \\for low temperature Ising model}

\author{
Aldo Procacci\inst{1}\and
Benedetto Scoppola\inst{2} \and
Elisabetta Scoppola\inst{3}}

%\date{}

\maketitle

\begin{center}
{\footnotesize
\vspace{0.3cm} \inst{1}  Departamento de Matem\'atica Instituto de Ci\^encias Exatas \\
Universidade Federal de Minas Gerais \\
Av. Ant\'onio Carlos, 6627 - Caixa Postal 702
30161-970 - Belo Horizonte - MG - BRASIL\\
\texttt{aldo@mat.ufmg.br}\\

\vspace{0.3cm}\inst{2}  Dipartimento di Matematica, Universit\`a di Roma
``Tor Vergata''\\
Via della Ricerca Scientifica - 00133 Roma, Italy\\
\texttt{scoppola@mat.uniroma2.it}\\

\vspace{0.3cm} \inst{3} Dipartimento di Matematica e Fisica, Universit\`a
Roma Tre\\
Largo San Murialdo, 1 - 00146 Roma, Italy\\
\texttt{scoppola@mat.uniroma3.it}\\ }

\end{center}

\vskip3.truecm

\begin{abstract}
We construct a parallel stochastic dynamics with invariant measure converging
to the Gibbs measure of the low temperature Ising model.
The proof of such convergence requires a polymer expansion based on suitably
defined Peierls-type contours.
\end{abstract}

\eject

%\tableofcontents
%%%%%%%%%%%%%SECT 1%%%%%%%%%%%%%%%%%%%%%%%%%
\section{Introduction}
\label{Intro}
\vspace{0.3cm}

 Goal of the paper is to construct a parallel stochastic dynamics
with invariant measure converging, in the thermodynamic limit, to the Gibbs measure
of the Ising model at low temperature.

The parallel dynamics described in this paper is a  homogeneous
discrete time Markov chain on a product space $S^\L=:\cX$, { where
$S=\{-1,1\}$ and $\L$ is a finite subset of $\Z^d$}, with transition
probabilities: \be{PCA} P(\s,\t)=\prod_{i\in \L}P(\t_i|\s). \ee

This kind of dynamics are known in the literature as Probabilistic Cellular Automata (PCA).

Parallel dynamics is a challenging {  topic} in Markov Chain Monte
Carlo (MCMC) methods and
 statistical mechanics
because they are
very promising  algorithms and  the efficiency of parallel computing
can be exploited in their simulation.
From a theoretical point of view there are few results in the literature on the
 convergence to equilibrium of parallel dynamics. This is  in general  a  difficult task, since the hight
mobility related to  parallelization, implies that it is much more complicated to individuate
possible bottleneck or saddle configuration in the tunneling between different configurations
on which  the invariant measure can be concentrated than  in the single spin flip dynamics.
We mention here two example where the
efficiency of parallel dynamics has been proved to be clearly higher than the efficiency of single spin flip dynamics.
The first
example is given in  \cite{dss2} where a control of the mixing time of an irreversible PCA related to the 2d Ising model
is given in a  particular regime of low temperature in a finite box of side $L$ with periodic boundary conditions. In this case
the mixing time turns out to be polynomial in $L$.
A second example is the Swendsen-Wang dynamics on Ising model \cite{SW} (see also the review \cite{M}). In this case the updating rule is not given
in the form (\ref{PCA}) of a PCA, but there is fast mixing because it is possible to update in a single
step of the Markov chain a large amount of spins.

However there is a preliminary more difficult  problem in the use of parallel dynamics in MCMC or in statistical mechanics:
 the control of their invariant measure.
This is the reason why their use is not widespread, even if PCA have been introduced in Equilibrium Statistical Mechanics a long time ago, see for instance \cite{LMS} and \cite{GKLM}.
 Invariant measures for infinite-volume PCA's may be non-Gibbsian (\cite{FT}) and even in the
finite volume case it is usually completely different w.r.t. the
invariant measure of Markov chain obtained by the random updating of
a single site $i$,  with the same  probability $P(\t_i|\s)$ used in (\ref{PCA}).
%i.e.,
%%$$P(\s,\t)=\frac{1}{|\L|}P(\t_i|\s)\qquad \hbox{ with }\qquad \t_j=\s_j\quad \forall
%%j\not=i$$
%%where $|A|$ denotes the cardinality of the set $A$}.
%%
%%
%%
%%{\bf \blu If we are supposing since the beginning $|S|=2$, i.e.  the spin at a site $i$ has only two possible values, then  maybe we could repeat what written in ref. \cite{dss2}}:
%%
%%{\red .... obtained by the random updating
%%of a single site $i$,  with the same  probability $P(\t_i|\s)$. More precisely,
%%
%$$
%\blu
%P(\s,\t)= \cases{{1\over |\L|}P(\t_i|\s) &if $\t=\s^i$\cr
%1- {1\over |\L|}P(\t_i|\s) &if $\t=\s$\cr
%0 & otherwise}
%$$
%{\red where $\s^i$ denotes the configuration obtained from $\s$ by
%flipping the spin at site $i$ and $|\L|$ denotes the cardinality of
%the set $\L$}.

In a previous paper \cite{dss1}, given a quite general spin system, a PCA was introduced with
 invariant measure converging, in the thermodynamic limit, to the Gibbs measure
of the Ising model at high temperature.
In the present paper we extend this result to the low temperature case.

We use here the same  construction of the parallel dynamics introduced in
\cite{dss1} (see also \cite{ISS}, { \cite{LS}} and \cite{GSSV}). We recall here the main ingredients.

%%%%%%%%
Given a spin configuration $ \s \in \cX$, % %\st{ where $S=\{-1,1\}$ and $\L$ is a finite subset of $\Z^d$},
 start with a Hamiltonian of
the form
\[
H(\s) := -\sum_{i,j} J_{ij} \s_i \s_j,
\]
corresponding to the Gibbs measure
\[
\pi_G(\s) %\propto \exp[-H(\s)]
{ =\frac{e^{-H(\s)}}{\sum_\s e^{-H(\s)}}}.
\]
This Hamiltonian can be {\em lifted} to a Hamiltonian on $ \cX\times \cX$, setting
\be{hst}
H(\s,\t) :=  -\sum_{i,j} J_{ij} \s_i \t_j + q\sum_i (1- \s_i \t_i).
\ee
{with the property $H(\s,\s)=H(\s)$} and
the PCA dynamics can be defined by
\be{P1}
P_{PCA}(\s,\t)=\frac{e^{-H(\s,\t)}}{\sum_\t e^{-H(\s,\t)}}
\ee
In the symmetric case $J_{ij}=J_{ji}$ {(and thus $H(\s,\t)=H(\t,\s)$)} it is immediate to prove that this is a reversible
PCA  with invariant measure
\be{P2}
\pi_{PCA}(\s)=\frac{\sum_\t e^{-H(\s,\t)}}{\sum_{\t,\t'} e^{-H(\t,\t')}}
\ee
%{\red
%In the antisymmetric case, i.e.  $J_{ij}\neq J_{ji}$ and thus ($H(\s,\t)\neq H(\t,\s)$) expression above for the invariant measure generally does not  hold.
%There are however particular  cases in which formula (\ref{piPCA}) for the invariant
% measure  when $H(\s,\t)\neq H(\t,\s)$  (see ahead).}

%%%%
%{\red
% In the  irreversible case, i.e. when $H(\s,\t)\not=H(\t,\s)$, the measure (\ref{P2}) is still the invariant measure of the Markov chain (\ref{P1})
% provided that
% $\sum_{\tau \in \cX} e^{-H(\s,\tau)} = \sum_{\tau \in \cX} e^{-H(\tau,\s)}$ (see Proposition 2.2. in \cite{LS}),
%which can be satisfied with a suitably choice of the non symmetric pair Hamiltonian (see \cite{dss2}, \cite{LS} and formula (\ref{ham2bis3}) ahead) and using periodic boundary conditions.
%}

We note that pair hamiltonians were already present in the
literature but their use was essentially related to generalizations
of the detail balance condition. See for instance in { \cite{OV}}
the notion of  ``approximately reversible non
degenerate" Markov chain. In these cases a pair hamiltonian was
introduced in order to write
$$
P(\s,\t)\propto e^{-[H(\s,\t)-H(\s)]}
$$
so that reversibility w.r.t. the Gibbs measure was immediately related to the
symmetry condition $H(\s,\t)=H(\t,\s)$.

Here the pair hamiltonian  {$H(\s,\t)$} has a different role since it is the necessary ingredient to define the dynamics.
More precisely,
to define the dynamics we want to
consider a { Gibbs} measure  $\m(\s,\t)\propto e^{-H(\s,\t)}$   on the space  of
 pairs of configurations $ (\s,\t)\in \cX^2$ instead of the Gibbs measure for
single configurations, since pairs of configurations
are possible moves of the dynamics.
% { Then the transition probability
%$P_{PCA}(\s,\t)$ turns out to be the conditional measure $\m(\s,\t)/\nu(\s)$ and the invariant measure
%$\pi_{PCA}(\s)$ turns out to be the marginal $\nu(\s)$ of the measure $\m(\s,\t)$}.
The ``lifting" given by the definition of pair Hamiltonian
(\ref{hst}) is due to the quadratic form of $H(\s)$ so that it is possible to consider
 $H(\s)=H(\s,\s)$.

%{\gre
 Once the pair hamiltonian is given, it is natural to define
 the measure on  $ \cX\times  \cX$:
$$\m(\s,\t)=\frac{e^{-H(\s,\t)}}{Z} \qquad\hbox{ with } \quad Z=\sum_{\s,\t}e^{-H(\s,\t)}.$$
The marginal of $\m(\s,\t)$ is the measure on $ \cX$:
$$
\sum_\t\frac{e^{-H(\s,\t)}}{Z}=:\nu(\s)\equiv \frac{Z_\s}{Z}
$$
and if $H(\s,\t)=H(\t,\s)$, i.e., in the {\it reversible case},  we have $\m(\s,\t)=\m(\t,\s)= \m(\{\s,\t\})$ and  $\nu(\s)=\sum_\t\frac{e^{-H(\t,\s)}}{Z}$.

For a given $\s\in \cX$ the transition probability of the chain associated to the measure
$\m$ is given by
\be{defdin}
P(\s,\t)=P((\s,\t)|\s)=\frac{\m(\s,\t)}{\nu(\s)}=\frac{e^{-H(\s,\t)}}{Z_\s}\equiv P_{PCA}(\s,\t)
\ee
and $\nu(\s)= \frac{Z_\s}{Z}=\pi_{PCA}(\s)$ is clearly its invariant measure.

The {\it irreversible case}, when $H(\s,\t)\not=H(\t,\s)$, is less trivial.
By considering periodic boundary condition, in the completely asymmetric case, see (\ref{ham2bis3}),  is possible to prove that
$$\nu(\s):=\sum_\t\frac{e^{-H(\s,\t)}}{Z}=\sum_\t\frac{e^{-H(\t,\s)}}{Z}$$
so that again $\nu$ is the marginal measure and equation (\ref{defdin})
  is again the natural definition of the dynamics.
%  }
%$$.$$

%%%%%%%%%%%%%%%%%%%%%%%%%%%%%%%%%%%%%%%%%%%%%%%%%%%%%%%%%%%%%
%{\bf Dubbio sul paragrafo in verde qui sopra: non mi \`e chiaro perch\`e abbiamo bisogno di definire la misura  $\m(\s,\t)$ su  $ \cX\times  \cX$. Abbiamo definito
%la catena di Markov (\ref{P1}), che \`e immediato provare che \'e reversibile  con  misura invariante su $S^\L$ data da (\ref{P2}) (basta verificare che $\pi_{PCA}P_{PCA}=\pi_{PCA}$).
%Non \`e tutto quello che ci serve per provare il teorema 1.1? Perch\`e  dobbiamo  anche far vedere che  la  misura invariante (\ref{P2})
%\`e la marginale della misura $\m(\s,\t)$?
%\bf Anche riguardo al caso irreversibile:
%se uno usa la Hamiltoniana non simmetrica $H^{I}(\s,\t)$
% per defire la catena di Markov (\ref{P1}),
%basta dire che, grazie alla  uguaglianza (\ref{sumsum}) provata nella prop. 2.1 in \cite{dss2},
%la catena di Markov ci ha ancora (\ref{P2}) come misura invariante. Se vale (\ref{sumsum}), allora uno pu\`o di nuovo verificare
% $\pi_{PCA}^I P_{PCA}^I=\pi_{PCA}^I $ senza bisogno di parlare di misure marginali su $S^\L\times S^\L$.
%
%La mia proposta \`e eliminare tutto il paragrafo verde e sostituirlo con la  frase in rosso  messa subito dopo la formula (\ref{P2}) (ma mi rimetto alla maggioranza):}

%%%%%%%%%%%%%%%%%%%%%%%%%%%%%%%%%%%%%%%%%%%%%%%%%%%%%%%%%%%%%%%%%%%%%%%%%%%%%%%%%%%%%%%%%%%%%%%%%%%

Note that the same construction holds here in the reversible and non-reversible case,
at least in { the} case of periodic boundary conditions.
This is an interesting feature of our approach, introducing
a unique promising language to
treat  equilibrium and non-equilibrium
statistical mechanics.
%%%%
\bigskip

 Define  the total variation distance, or $L_1$ distance,
between $\pi_G$ and $\pi_{PCA}$ as
\be{dist}
\| \pi_{PCA}-\pi_{G}\|_{TV}=\frac{1}{2}\sum_{\s\in \cX}|\pi_{PCA}(\s)-\pi_{G}(\s)|
\ee
The parameter $q$  controls
 the average number of spin-flips in a single step of the dynamics.
It was proved in \cite{dss1} that,  defining
$\d:=e^{-2q}$, if
$\d = \d(|\L|)$ is such that $\lim_{|\L|\to\infty}\d^2|\L|=0$ and if the temperature is sufficiently
high  (i.e. $J_{i,j}$ sufficiently small) then
\be{conv}
\lim_{|\L|\to\infty}\| \pi_{PCA}-\pi_{G}\|_{TV}=0.
\ee

%%%
In the present paper we use completely different tools to prove a similar result
in the low temperature regime. Indeed we will use
a polymer expansion based on suitably defined Peierls-type contours.
The   cluster expansion techniques involved in the present paper
are quite robust. We present the results in the simple case
of two dimensions:
\begin{itemize}
\item
with plus boundary conditions in the reversible case;
\item
with periodic boundary conditions in the reversible case;
\item
with periodic boundary conditions in the irreversible case.
\end{itemize}
As far as the choice of parameters is concerned, we are in a quite general regime.
We  need again the hypothesis $\lim_{|\L|\to\infty}\d^2|\L|=0$ but  the low temperature
regime is not related to a particular asymptotic in the limit $|\L|\to\infty$.
In the irreversible case a similar result was obtained in \cite{dss2} but in a very particular regime of
low temperature, where inverse temperature suitably increases with the volume.

%{Note}
{ We finally remark} that in this paper
estimates are not optimized.

%As noted above the interest of such a result is quite evident from a simulation point of view,
%as discussed in section ***???

%\bi
%\item {\bf descrizione risultato}
%\item {\bf relazione con  [dss1],  qui bassa temperatura ma interazione di Ising,  relazione con [dss2], qui caso reversibile, caso  generale di $h$ e bc in cui non abbiamo stima convergenza, problematica generale}
%\item{\bf perche' questo lavoro e' interessante}
%\item {\bf strumento diverso: cluster expansion}
%\item {\bf possibili generalizzazioni e estensioni}
%\ei
%*****

%%%%%%%%%%%%
\subsection{Definitions}\label{definitios}
\label{def}
{ Henceforth $\L$ denotes a two-dimensional $L\times L$ square lattice in $\mathbb{Z}^2$   and
$B_\L$  denotes the set of all nearest neighbors
in $\L$, i.e. $B_\L=\{\langle i,j\rangle:\;i,j\in\L,\; |i-j|=1\}$ with $|i-j|$ being the usual lattice distance in $\mathbb{Z}^d$.
% and $|B_\L|=2L(L-1)$.
We denote by $\partial^{ext}\L$  ($\partial^{int}\L$)  the external (internal) boundary of $\L$, i.e.
$\partial^{ext}\L=\{i\in \mathbb{Z}^2\setminus\L:\; \exists j\in \L \,\,{\rm s.t.}\,\,|i-j|=1\}$   ($\partial^{int}\L=\{i\in \L :\; \exists j\in \mathbb{Z}^2\setminus\L \,\,{\rm s.t.}\,\,|i-j|=1\}$)
and we  set $\bar \L=\L\cup \partial^{ext}\L$

We set $B^{per}_\L=\{\<i,j\>: \; i,j\in\L\;{\rm and}\; {\rm either}\;|i-j|=1\;{\rm or}\; |i-j|=L-1\}$. Namely, $B^{per}_\L$ is the set of all nearest neighbors
in $\L$ plus the pairs of sites at
opposite faces of the square $\L$,
% (therefore $|B^{per}_\L|=2L^2$).
so that  the pair $(\L, B^{per}_\L)$ is homeomorphic to the two-dimensional discrete torus $(\mathbb{Z}/(L\mathbb{Z})^2$.
%. %($B^+_\L=\{\{i,j\}\subset \bar\L:\; |i-j|=1\}$ ).

We set $B^{+}_\L=\{\<i,j\>:\;i,j\in \bar \L:\; |i-j|=1\}$, i.e. $B^{+}_\L$ is the set of all nearest neighbors in $\bar\L$.
%(therefore $|B^{per}_\L|=2L(L+1)$).
%Let us denote shortly $\cX^+$  the set of spin configurations with $+$ boundary conditions (b.c.), i.e., the set of functions
%$\s:\bar\L\to\{-1,1\}^{|\bar\L |}$ such that $\s_i=+1$ if $i\in\partial^{ext}\L$.
%
We finally recall that  $\cX$ denotes the set of spin configurations in $\L$., i.e., $\cX$ is the set of all functions
$\s:\L\to\{-1,1\}$.}

In both cases of  periodic and $+$  b.c. define the Ising hamiltonian with interaction $J>0$ { as follows. Given $\s\in \cX$},
\be{ham}
H^{per}(\s)=-J\sum_{\langle i,j\rangle\in B_{\L}^{per}} \s_i\s_j,\qquad
H^{+}(\s)=-J\sum_{\langle i,j\rangle\in B_{\L}^{+}} \s_i\s_j
\ee
{ where by convention $\s_i=+1 $ when $i\in \partial^{ext}\L$}.

%where $B_{\L}^{per}$ is the set of nearest neighbor bonds in the torus $\L$. {while $B_{\L}^+$ is the set of nearest neighbour bonds in $\L$ including also the bonds
%$\langle i,j\rangle$ with $i\in\L,\; j\in \partial^{ext}\L$}.

\noindent
{ Using the notation ${H^*(\s)}$ with either $*=per$ or $*=+$, we denote
\be{wG}
w_G^{*}(\s)=e^{-H^{*}(\s)},~~~~~~~~ Z_G^*=\sum_{\s\in \cX}e^{-H^{*}(\s)}
\ee
so that  }   the standard Gibbs measure with $*=per,+$  is
\be{gibbs}
\pi_G^{*}(\s)=\frac{w_G^{*}(\s)}{Z_G^{*}}%=:\frac{w_G^{*}(\s)}{\sum_{\s\in\cX}w_G^{*}(\s)}%, \hskip 1cm
%\pi_G^+(\s)=\frac{e^{-H^+(\s)}}{Z_G^+}=:\frac{w_G^+(\s)}{\sum_{\s\in\cX^+}w_G^+(\s)}
\ee

We now define, for every pair of configurations $\s,\t\in \cX$   the following pair Hamiltonians { for periodic and $+$  b.c. respectively}:

\be{ham2}
H^{per}(\s,\t)=-\frac{J}{2}\sum_{i\in\Lambda} \sum_{j\in \L\atop \langle i,j\rangle\in B_{\L}^{per}}\s_i\t_j+q\sum_{i\in\Lambda}(1-\s_i\t_i)
\ee
\be{ham2+}
H^+(\s,\t)=-\frac{J}{2}\sum_{i\in\Lambda} \sum_{\stackrel{j\in\Lambda:}{\langle i,j\rangle\in B_{\L}}}\s_i\t_j-
\frac{J}{2}\sum_{i\in\partial^{int}\Lambda}\sum_{\stackrel{ j\in\partial^{ext}\Lambda:}
{|i-j|=1}}
(\s_i+\t_i)+q\sum_{i\in\Lambda}(1-\s_i\t_i)
\ee
where %{\bf   $\partial^{int}\L$ is the internal boundary of $\L$} and $|i-j|$ denotes the usual lattice distance and
$q>0$ is a 
volume-dependent 
parameter that we will
choose later  (see Theorem \ref{t1} ahead).
Note that $\sum_{\stackrel{ j\in\partial^{ext}\Lambda:}
{|i-j|=1}}
$
is equal to one in each site in $\partial^{int}\Lambda$ which is not a corner
and is equal to 2 in the four corners of $\partial^{int}\Lambda$.

\noindent
{
Clearly,  for $*=+,per$, $H^*(\s,\t)$ is symmetric, i.e.} $H^*(\s,\t)=H^*(\t,\s)$. { Moreover observing that $H^{per}(\s)$ and $H^{+}(\s)$ can be rewritten respectively as}
\be{hamper}
H^{per}(\s)%=-J\sum_{\langle i,j\rangle\in B_{\L}^{per}} \s_i\s_j
= -{J\over 2}\sum_{i\in \L}\sum_{j\in \L\atop\langle i,j\rangle\in B_{\L}^{per}}\s_i\s_j
\ee
\be{ham+1}
H^{+}(\s)%=-J\sum_{\langle i,j\rangle\in B_{\L}} \s_i\s_j-J\sum_{i\in \partial^{int}\L}\sum_{j\in \partial^{ext}\L\atop |i-j|=1}\s_i
=
-{J\over 2}\sum_{i\in \L}\sum_{j\in \L\atop \langle i,j\rangle\in B_{\L}}\s_i\s_j-J\sum_{i\in \partial^{int}\L}\sum_{j\in \partial^{ext}\L\atop |i-j|=1}\s_i
\ee
{ we also get immediately that} $H^*(\s,\s)=H^*(\s)$.

%\be{ham1}
%\cancel{H^*(\s,\s)=H^*(\s)\qquad \hbox{ and }\qquad H^*(\s,\t)=H^*(\t,\s)}, %\hskip 1cm {\rm with}\;\;*=per,+
%\ee

\noindent
{
Let us write
\be{defh2}
\partial_i=
\frac{J}{2}{\bf 1}_{\{i\in\partial^{int}\Lambda\}}\sum_{\stackrel{ j\in\partial^{ext}\Lambda:}
{|i-j|=1}}1,\qquad  G(\s)=\sum_{i\in\Lambda}\partial_i\s_i
\ee
and define }
\be{defh}
h_i^+(\s)= \frac{J}{2}\sum_{\stackrel{j\in\L:}{\langle i,j\rangle\in B_{\L}}}\s_j+\partial_i,
\qquad
h_i^{per}(\s)= \frac{J}{2}\sum_{\stackrel{j\in\L:}{\langle i,j\rangle\in B^{per}_{\L}}}\s_j.
\ee
Then we can rewrite
\be{ham2bis2}
H^*(\s,\t)=-\sum_{i\in\Lambda}(h^*_i(\s)+\s_iq)\t_i -G^*(\s)+q|\L|
\ee
{
with
$G^+(\s)=G(\s)$ and $G^{per}(\s)=0$.}
%Setting now

%\be{htil}
%\Ht^*(\s,\t)= -\sum_{i\in\Lambda}(h^*_i(\s)+\s_iq)\t_i -G^*(\s)
%\ee
We can { now } define  parallel dynamics, that we will call
{\it PCA dynamics}, with the following
transition probabilities, for $*=per,+$
\be{pss}
P_{PCA}^*(\s,\t)=\frac{e^{-H^*(\s,\t)}}{\sum_{\t\in\cX}e^{-H^*(\s,\t)}}
%
%=\frac{e^{-\Ht^*(\s,\t)}}{\sum_{\t\in \cX}e^{-\Ht^*(\s,\t)}}
\ee
%\st{where $\cX^{per}\equiv \cX$.}
It is a standard task to show that this PCA dynamics is reversible with respect to the measure
\be{ps}
\pi_{PCA}^*(\s)=\frac{\sum_{\t\in \cX} e^{-H^*(\s,\t)}}{\sum_{\t,\t'\in \cX} e^{-H^*(\t,\t')}}
%=\frac{\sum_{\t\in \cX} e^{-\Ht^*(\s,\t)}}{\sum_{\t,\t'\in \cX} e^{-\Ht^*(\t,\t')}}
\ee

\noindent
Due to   {  (\ref{ham2bis2})},
the transition probabilities of this Markov chain
can be written as a product of the transition probabilities of each component  $\t_i$ of the new configuration $\t$,
as usual for PCAs:
$$
P^{*}_{PCA}(\s,\t)=\prod_{i\in \L}P^{*}(\t_i|\s)
$$
with
$$P^{*}(\t_i|\s)
=\frac{e^{(h_i^{*}(\s)+\s_iq)\t_i}}
{2\cosh(h_i^{*}(\s)+\s_iq)}.
$$
%\st{and similarly in the case of $+$ b.c.}

In the case of periodic boundary condition we will also consider an {\it irreversible} parallel dynamics, denoted with
$I$ for irreversible, %{ \st{(but omitting $^{per}$ for  periodic b.c.),}}
by considering the following
pair hamiltonian
\be{ham2bis3}
%\begin{split}
H^{I}(\s,\t)=
 - \sum_{i \in \L} \left[J \s_i(\t_{i^{\uparrow}} + \t_{i^{\rightarrow}}) +q\s_i \t_i\right]   = - \sum_{i \in \L}\left[J \t_i(\s_{i^{\downarrow}} + \s_{i^{\leftarrow}})  +q\s_i \t_i\right]
%\end{split}
\ee
where $i^{\uparrow}, i^{\rightarrow}, i^{\downarrow}, i^{\leftarrow}$ are respectively the up, right, down, left neighbors of the site $i$ on the torus $(\L,B^{per}_\L)$.
{ We have $H^{I}(\s,\s)=H^{per}(\s)-q|\L|$} but
note that now
$H^{I}(\s,\t)\not=H^{I}(\t,\s)$.  However, as shown in { \cite{dss2, LS}},  the following {\em weak symmetry condition (dynamical balance)} holds
\be{sumsum}
\sum_{\tau \in \cX} e^{-H^{I}(\s,\tau)} = \sum_{\tau \in \cX} e^{-H^{I}(\tau,\s)} .
\ee
so that the parallel dynamics defined by
$$
P^I_{PCA}(\s,\t)=\prod_{i\in \L}\frac{\exp\left\{ \tau_i\left[J(\s_{i^{\downarrow}} + \s_{i^{\leftarrow}}) + q \s_i\right]\right\}}{2 \cosh(J(\s_{i^{\downarrow}} + \s_{i^{\leftarrow}}) + q \s_i)}
$$
is irreversible with a unique stationary distribution $\pi^I_{PCA}$ given by
\[
\pi^I_{PCA}(\s) := \frac{Z^I_{\s}}{Z^I_{PCA}},\qquad Z^I_{\s}=\sum_{\t\in\cX}e^{-H^{I}(\s,\t)}
\]
with $Z^I_{PCA} := \sum_{\s} Z^I_{\s}$.

%%%%%%%%%%

\noindent
\subsection{Results}
\label{result}

\noindent
By using the previous definitions we can 
now
 state our general results.
The first theorem states that the stationary
measure of the reversible PCA and the Gibbs measure
are  equivalent in the thermodynamical limit.

\bt{t1}
{
Set $\d=e^{-2q}$, and let $\d$ be such that
\be{conddelta}
\lim_{|\L|\to\infty}\d^2|\L|=0,
\ee
}
then, there exist $J_c$  such that  for any $J>J_c$
\be{th1.1}
\lim_{|\L|\to\infty}\| \pi^*_{PCA}-\pi^*_{G}\|_{TV}=0\qquad *=+,per
\ee

\et

\noindent
The second theorem deals with the irreversible case, with periodic boundary conditions

\bt{t2}

Under the assumption (\ref{conddelta}), there exist $J_c$  such that  for any $J>J_c$
\be{th2}
\lim_{|\L|\to\infty}\| \pi^{I}_{PCA}-\pi^{per}_{G}\|_{TV}=0
\ee

\et

%%%%

\br{r}
Let us make some comments on the results and its relations with the similar
results obtained in \cite{dss1} and \cite{dss2}.
\bi
\item[i)]
First note that (\ref{conddelta}) corresponds to say that the parameter $q$ goes to infinity faster than
$\frac{1}{ 4}\ln|\L|$. This hypothesis was also assumed in  \cite{dss1}.
\item[ii)] As far as Theorem \ref{t1} is concerned, as mentioned in the introduction,
a similar result was obtained  in  \cite{dss1} but in the opposite regime of high temperature (uniqueness of phase for the Gibbs measure). 
\item[iii)]
As far as Theorem \ref{t2} is concerned a similar result was obtained  in  \cite{dss2}
but with a very particular choice of the parameters. In that paper indeed
a low temperature regime was defined by fixing $J(L)$ and $q(L)$ as function
of the side $L$ of the square $\L$. In the present paper we want to stress that
we are in a true low temperature regime, i.e., we prove the convergence of the invariant measures of the PCA to the Gibbs measure in the thermodynamical limit for any $J$ 
large enough provided that the hypothesis (\ref{conddelta}) on the parameter $q$
is satisfied.

\item[iv)]
Actually the idea
of the proof of Theorems \ref{t1} and \ref{t2} follows the same line of the similar results in  \cite{dss1} and \cite{dss2}. The strategy is  to prove that the following inequality holds
\be{thclus}
\| \pi^*_{PCA}-\pi^*_{G}\|_{TV}={\cal O}(\d |\L|^{1/2})
\ee
and then immediately concluding the proof by using the hypothesis (\ref{conddelta}).
However the proof of (\ref{thclus}), is based on  Lemmata \ref{main}
and \ref{mainirr} which are completely different w.r.t. the tools used in the previous papers  \cite{dss1}, \cite{dss2}.
\ei
\er

\noindent

%%%%%%%%%%%%%%%%%%

\section{Proof of Theorems \ref{t1} and \ref{t2}}
\label{proof}
\noindent
Let us start with Theorem \ref{t1}.
We first prove that for any fixed $J>J_c$
there exist $\d_J$ such that
for any $\d<\d_J$  (\ref{thclus}) holds.

\noindent
{ Let, for $*=per,+$
\be{wpca}
w^*_{PCA}(\s)=\sum_{\t\in \cX}e^{-H^*(\s,\t)}
\ee
Then,
recalling (\ref{ham2bis2}), we have (modulo the constant 
$e^{-q|\L|}$
 which cancels out in the ratio $w^*_{PCA}(\s)/\sum_\s w^*_{PCA}(\s)$)}
$$
w^*_{PCA}(\s)=e^{G^*(\s)}\sum_{\t\in\cX} e^{\sum_{i\in \L}[(h^*_i(\s)+q\s_i)\t_i)]}.
$$
%$$
%w^{per}_{PCA}(\s)=\sum_{\t\in\cX^{per}} e^{\sum_{i\in \L}[(h_i^{per}(\s)+q\s_i)\t_i)]}.
%$$
We now rewrite the sum on $ \t\in\cX$ in the following way. Given   $ \s\in\cX$,
to sum over all $ \t\in\cX$   is the same as to sum over
all subset $I\subset\L$ such that $\t_i=-\s_i$ if $i\in I$ while $\t_i=\s_i$ otherwise.
Hence we can write
$$
w^*_{PCA}(\s)=e^{G^*(\s)}\sum_{I\subset \L}e^{\sum_{i\in \L}h^*_i(\s)\s_i-2\sum_{i\in I}h^*_i(\s)\s_i-2q|I|}=
$$
\be{ww}
=e^{\sum_{i\in \L}h^*_i(\s)\s_i+ G^*(\s)}
\sum_{I\subset \L}\prod_{i\in I}e^{-2h^*_i(\s)\s_i-2q}=
e^{\sum_{i\in \L}h^*_i(\s)\s_i+ G^*(\s)}\prod_{i\in \L}(1+\delta\phi^*_i)
\ee
where,  recalling  that $\d=e^{-2q}$,  we have defined
\be{phi*}
\phi^*_i=e^{-2(h^*_i(\s))\sigma_i}
\ee
{ By definitions (\ref{hamper}), (\ref{ham+1}), (\ref{defh2}) and (\ref{defh}) it is easy to check that}
$$
\sum_{i\in \L}h^*_i(\s)\s_i+ G^*(\s)=-H^*(\s)
$$
{ Therefore, recalling (\ref{wG}) and setting}
\be{f}
 f^*(\sigma)=\prod_{i\in \L}(1+\delta\phi^*_i),
\ee
{ we get}
\be{wpcsf}
w^*_{PCA}(\s)= w^*_{G}{ (\s)}f^*(\sigma)
\ee
%For periodic boundary condition the same argument gives
%\be{ww2}
%w^{per}_{PCA}(\s)=e^{\sum_{i\in \L}h_i^{per}(\s)\s_i}
%\sum_{I\subset \L}\prod_{i\in I}e^{-2h_i^{per}(\s)\s_i-2q}=
%w^{per}_{G}(\s)\prod_{i\in \L}(1+\delta\phi^{per}_i)
%\ee
%where
%$$\phi^{per}_i=e^{-2h_i^{per}(\s)\sigma_i}$$
{ and  therefore, by definitions (\ref{wpca}), (\ref{gibbs}) and (\ref{ps})} we have
$$
\pi^*_{PCA}(\s)={w^*_{PCA}(\s)\over \sum_{\s\in\cX} w^*_{PCA}(\s)}={w^*_{G}(\s)f^*(\sigma)\over \sum_{\s\in\cX} w^*_{G}(\s)f^*(\sigma)}
={{w^*_{G}(\s)\over Z^*_G}f^*(\sigma)\over \sum_{\s\in\cX} {w^*_{G}(\s)\over Z_G^*}f^*(\sigma)}=
{\p^*_G(\s)f^*(\s)\over \p^*_G(f^*)}
$$
%\be{zz}
%\cancel{Z^*_{PCA}=\sum_\s w^*_{G}(\s)f^*(\sigma)=Z^*_{G}\pi^*_G(f^*).}
%\ee
Hence
%$$
%\cancel{\| \pi^*_{PCA}-\pi^*_{G}\|_{TV}=\sum_\s \frac{w^*_{G}(\s)}{Z^*_G}\left| \frac{\pi^*_{PCA}(\s)}{\pi^*_{G}(\s)}-1\right|
%=\sum_\s \frac{w^*_{G}(\s)}{Z^*_G}\left| \frac{w^*_{PCA}(\s)}{w^*_{G}(\s)}\,\frac{Z^*_{G}}{Z^*_{PCA}}-1\right|}
%$$
\be{stid}
\| \pi^*_{PCA}-\pi^*_{G}\|_{TV}=\sum_\s \p^*_G(\s)\left| \frac{f^*(\s)}{\pi^*_{G}(f^*)}-1\right|=
\pi^*_{G}\left(\left| \frac{f^*(\s)}{\pi^*_{G}(f^*)}-1\right|\right)\le\frac{({\rm{var}}_{\pi^*_G}(f^*))^{1/2}}{\pi^*_G(f^*)}
=\big(\D^{ *}(\d)\big)^{1/2}\ee
with

\be{Del}
\D^*(\d)=\frac{\pi_G((f^{*})^2)}{(\pi_G(f^*))^2}-1
\ee
By writing
$$
\D^*(\d)=\exp \Big[\ln \pi^*_G((f^*)^2)-2 \ln \pi^G(f^*)\Big] -1
$$
 Theorem \ref{t1} is proved by the following:
\bl{main}
There exists $J_c$ such that for  any $J>J_c$
\bi
\item[i)] \hskip 1cm
$\displaystyle\frac{\ln \p^*_G((f^*)^2)}{|\L |}$ ~and~ $\displaystyle\frac{\ln \p^*_G(f^*)}{|\L |}$ are analytical functions of $\d$
for $|\d|<\d_J$.
\item[ii)] \hskip 1cm
$\displaystyle\frac{\ln \p^*_G((f^*)^2)}{|\L |}-2\frac{\ln \p^*_G(f^*)}{|\L |}=O(\d^2)$
\ei
\el
%%%
\noindent
{\bf Proof}

\noindent
{\it Part i)}

\noindent
We will show the analyticity of $\frac{\ln \p^*_G((f^*)^2)}{|\L |}$ and $\frac{\ln \p^*_G(f^*)}{|\L |}$
by showing that both these quantities may be written as partition functions of an
abstract polymer gas. Then the analiticity will follow by standard cluster expansion, see \cite{
kp, 
BFP,fp}.

\noindent
{ We have
\be{piG}
\pi^*_G((f^{*})^k)=\frac{1}{Z^+_G}{\sum_{\s\in \cX} e^{-H^*(\s)}\prod_{i\in \L}(1+\delta\phi^*_i)^k},~~~~~~~~~~~~ k=1,2
\ee
and, recalling (\ref{defh}), (\ref{phi*}), note that
\be{hstari}
\phi^*_i=\exp\{-2h^*_i(\s)\s_i\}=\exp\{-J\sum_{j\in\L^*\atop \<i,j\>\in B^*_\L} \s_i\s_j\}
\ee
with $\L^+=\bar\L$, $\L^{per}=\L$ and with the usual convention that $\s_j=+1$ when $j\in \partial^{ext}_\L$}.

We first consider  $\p^+_G((f^+)^k),\  k=1,2$.
 Then, by (\ref{hstari}), we can write
\be{effekappa1}
\p^+_G((f^+)^k)=
\frac{1}{Z^+_G}\sum_{\s\in \cX}\exp\{J\sum\limits_{\langle i,j\rangle\in B^+_{\L}}\s_i\s_j\}
\prod_{i\in\L}\left(1+\d \exp\{-{ J}\sum\limits_{j\in\bar\L,|j-i|=1}\s_i\s_j\}\right)^k
\ee
{ with the convention that $\s_j=+1$ whenever $j\in\partial^{ext}\L$}.

\noindent

We rewrite the l.h.s. of (\ref{effekappa1}) via a standard
Peierls contour gas. Following the usual construction, for a fixed configuration
$\s$, let $\G$ be the set of unit segments perpendicular to the center of each
bond  of nearest neighbors in $\bar\L$ having opposite spins at its
extremes (again with the convention that $\s_i=1$ if $i\in \partial^{ext}\L$).
Any  unit segment  $e\in\G$ is a nearest neighbor bond of a $(L+1)\times(L+1)$ square with vertices in the dual unit square lattice  $\mathbb{Z^*}^2$
(translated by the vector $({1\over 2},{1\over 2})$  respect to the original lattice  $\mathbb{Z}^2$).
As far as   + b.c are concerned, the correspondence $\s\mapsto\G$ is one-to-one
and the unit segments of $\G$ form a collection of closed polygons
which separate regions where the spins are positive from regions
where they are negative. These polygons
possibly intercept themselves
in such a way that the degree of each vertex is even.
%only on vertices in such a way that their vertices belong to either two or four sides.
%, so that we may write $\G\subset B_{\tilde \L}$
%(viewing $B_{\tilde \L}$ as a set of $2L(L+1)$ unit segments, one for each nearest neighbor pair).
Let us denote by ${\cal G}_\L$ the set of all possible $\G$.
% (one $\G\in{\cal G}_\L $ for each $\s\in S^\L$).
%Let for a  given $\s\in S^\L$, $E^\pm_{ \L}=\{\<i,j\>\in B_{\bar\L}:\; \s_i\s_j=\pm 1\}$, and let
% $\G$ be the  set of unit  segments  uniquely associated to $\s\in S^\L$. Then $|\G|=|E^-_{\L}|$ and $|E^+_{\L}|+|E^-_{ \L}|=|B_{\bar\L}|$ and therefore we have
$$
 J\sum\limits_{\langle i,j\rangle\in B_{{\bar\L}}}\s_i\s_j~=
~
%J (|E^+_{ \L}|-|E^-_{ \L}|)=
|B_{\bar\L}|-2 J|\G|
$$
\noindent
Moreover,  calling $l_s(\G)$,  for  $s=1,...,4$,  the set of vertices $i\in\L$ having exactly
 $s$ edges $\langle i,j\rangle$ with dual in $\G$, %(i.e. having $s$ unit segments of $\G$ with midpoint at distance ${1/2}$),
 we have
$$
 \prod_{i\in\L}\left(1+\d \exp\{-{ J}\sum\limits_{j\in\bar\L,|j-i|=1}\s_i\s_j\}\right)^k= \prod_{i\in\L\setminus\cup_{s=1}^4 l_s(\G)}\left(1+\d e^{- 4J}\right)^k
\times\prod_{i\in  l_1(\G)}\left(1+\d e^{- 2J}\right)^k\times
$$
$$
 \times \prod_{i\in  l_2(\G)}\left(1+\d \right)^k\times \prod_{i\in  l_3(\G)}\left(1+\d e^{+  2J}\right)^k\times\prod_{i\in  l_4(\G)}\left(1+\d e^{+ 4J}\right)^k
$$
Therefore (\ref{effekappa1}) can be rewritten as follows
\be{effekappa2}
\p^+_G((f^+)^k)=\frac{1}{Z^+_G}e^{J|B_{\bar\L}|}(1+\d e^{- 4J})^{k|\L|}\Xi^{+(k)}_\L(J,\d)
\ee
where
\be{xi+}
\Xi^{+(k)}_\L(J,\d)=\sum_{\G\in {\cal G}_\L}
\left[e^{-2J|\G|}\xi_k(\G)\right]
\ee
with
\be{xi}
\xi_k(\G)=\left[\frac{1+\d e^{- 2J}}{1+\d e^{- 4J}}\right]^{k|l_1(\G)|}
\left[\frac{1+\d }{1+\d e^{- 4J}}\right]^{k|l_2(\G)|}
\left[\frac{1+\d e^{+ 2J}}{1+\d e^{- 4J}}\right]^{k|l_3(\G)|}
\left[\frac{1+\d e^{+ 4J}}{1+\d e^{- 4J}}\right]^{k|l_4(\G)|}
\ee
{ From (\ref{effekappa2}) we obtain that
$$
{1\over |\L|}\ln \p^+_G((f^+)^k)=-{\ln{Z^+_G}\over |\L|} + J{|B_{\bar\L}|\over |\L|}+k\ln(1+\d e^{- 4J})+ {1\over |\L|}\ln \Xi^{+(k)}_\L(J,\d)
$$
The first two terms of the l.h.s. of identity above do not depend on $\d$ and are uniformly bounded as $\L\to\infty$ and  the third term is analytic in $\d$ and independent of $\L$. It thus remains to
be analyzed the term ${1\over |\L|}\ln \Xi^{+(k)}_\L(J,\d)$}.
In order to do that, we regard each $\G\in {\cal G}_\L$ as the disjount union of its suitably defined
{
\it
p-connected components
} $\g_1,...,\g_n$
{ in such a way that any vertex of $l_s(\G),\ s=1,...,4$, is associated to at most one 
p-connected component $\g_i$}.
We introduce the notion of p-connection as an extension of the usual notion of connection. 
Namely we define
   two unit segments
of $\G$ as 
{\it p-connected}  if they share a common vertex {\it or if they are parallel and at distance 1}. 
 { We denote by ${\cal P}_{\L}$ the set of all such 
 p-connected components, i.e., $\g_i\in {\cal P}_{\L}$.
 For $\g,\g'\in {\cal P}_{\L}$ we write $\g\sim \g'$
 if $\g\cup\g'\not\in{\cal P}_{\L}$.
With these notations}  (\ref{xi+})  can be written as follows
\be{csi}
\Xi^{+(k)}_\L(J,\d)=\sum_{n\ge 0}{1\over n!}\sum_{(\g_1,...,\g_n)\in {\cal P}_\L^n\atop \g_i\sim \g_j}\prod_{i=1}^{n}
\left[e^{-2J|\g_i|}\xi_k(\g_i)\right]
\ee
where
the $n=0$ contribution is 1, and it
corresponds of course to the configuration $\s=1$.

The r.h.s. of (\ref{csi})
is the  grand canonical partition function of a hard-core polymer gas, where the polymers
are the elements of ${\cal P}_\L$ defined above and their activity $\r_k(\g)$ is
\be{acti}
\r_k(\g)=\xi_k(\g)e^{-2J|\g|}
\ee

\noindent
{ It is well known that the logarithm of $\Xi^{+(k)}_\L(J,\d)$  divided by $|\L|$
can be written as an absolutely convergent series uniformly in $\L$ for $\r_k(\g)$ sufficiently small. For the sake of simplicity, and since we are not seeking optimal 
bounds, we will use the Kotecky-Preiss (KP) condition \cite{kp} 
(see also \cite{BFP}, \cite{fp} for better conditions). 
In its general form, the KP condition reads
\be{kp1}
\sum_{\g\nsim\g'}\r(\g)e^{a(\g)}\le a(\g')
\ee
where $a(\g)$ is any positive function of $\g$.
Choosing $a(\g)=a|\g|$, $a>0$, and noting that 

\be{kp2}
\sum_{\g\nsim\g'}\r(\g)e^{a|\g|}\le 3|\g'|\sup_{x\in \mathbb{Z}^2}
\sum_{\g\ni x}\r(\g) e^{a|\g|}
\ee

we have that equation (\ref{kp1}) is satisfied if
\be{fp}
\sup_{x\in \L}\sum_{\g\in {\cal P}\atop x\in \g}\r_k(\g)e^{a|\g|}\le \frac{a}{3}
\ee
 where ${\cal P}$ is the set of all p-connected contours  in $\mathbb{Z}^2$.
The factor 3 in the r.h.s. of (\ref{kp2}) comes from the fact that the contour
$\g$ is anchored to $\g'$ if there exists a vertex $x$ of $\g'$ of degree 2 in $\g'$
such that either $x$ is also a vertex of $\g$ or the edge 
exiting from $x$ in $\g'$, when $\g'$ is traveled in a given
direction, is parallel and at distance 1 to an edge of $\g$.
In the latter case there are two possible choices, hence the factor 3.

{
% and
%$e$ is any fixed nearest neighbor bond in $\mathbb{Z}^2$ (due translational invariance of  $\r_k(\g)$).
%In general the parameter $a$ is chosen in such way to optimize bounds that result from inequality (\ref{fp}).
As we said in the introduction it is not our intention to look for optimal estimate. We thus
choose   $a=1$ (which is not optimal) and  hence  condition (\ref{fp}) becomes}
\be{fp2}
\sup_{x\in \mathbb Z^2}\sum_{\g\in {\cal P}\atop x\in \g}\r_k(\g)e^{|\g|}\le \frac{1}{3}
\ee
{ To check (\ref{fp2}), let first obtain an upper} bound for $\r_k(\g)$.  Recalling the definition (\ref{xi}), we get,
for $|\d|<e^{ 4J}$
\be{st1}
\r_k(\g)\le e^{-2J|\g|}\left[\frac{1+|\d| e^{+ 4J}}{1-|\d| e^{- 4J}}\right]^{k\sum_{s=1}^4|l_s(\g)|}
\ee
Observing that
$$
\sum_{s=1}^4|l_s(\g)|\le 2|\g|
$$
we get, { for any $k=1,2$}
\be{st2}
\r_k(\g)\le e^{-2J|\g|}\left[\frac{1+|\d| e^{+ 4J}}{1-|\d| e^{- 4J}}\right]^{2k|\g|}
\le
\left(e^{-2J}\left[\frac{1+|\d| e^{+ 4J}}{1-|\d| e^{- 4J}}\right]^{4}\right)^{|\g|}
\ee

%%%%%%
Let us  call
$$
A(J,\d)=e^{-2J}\left[\frac{1+|\d| e^{+ 4J}}{1-|\d| e^{- 4J}}\right]^{4}
$$
When the 
p-connected contour $\g\in{\cal P}_\L$ is composed by several different connected 
components we define a new connected contour by adding edges
between adjacent contours. We define a rule. For instance we order
the vertices of $\L$ and
we consider the consequent order of edges 
with lexicographic ordering.
Given a contour $\g$, we add to the contour 
 pairs of edges corresponding to
 the set of smallest edges
in order to obtain a connected graph. 
The resulting graph  $\bar\g$
is then a connected  graph  with possible multiple edges (the added pairs), with even
degree at each vertex, with maximal degree 4.
Since each connected
component of $\g$ has at least 4 edges, the number of added 
pairs of  edges is at most 
$\frac{|\g|}{4}$.

Then
$$
\sup_{x\in \mathbb{Z}^2}
\sum_{\g\ni x}\r_k(\g)e^{|\g|}\le \sum_{n\ge 4}A(J,\d)^n e^n
\sum_{\bar\g\ni 0\atop |\g|=n}1
\le
\sum_{n\ge 4}[e\cdot 3^{
3/2}A(J,\d)]^n
$$
Here the factor $3^{
3n/2}$ comes from the estimate of the number of graphs $\bar\g$ of
$
3n/
2$ edges passing through a fixed vertex. 
Indeed the graph $\bar\g$ has a Eulerian circuit 
and, as usual, each new step has three possible choices.
Therefore (\ref{fp2}) is surely fulfilled  for $k=1,2$  if
\be{akj}
{[e\cdot 3^{3/2}A(J,\d)]^4\over 1-e\cdot 3^{3/2}A(J,\d)}\le \frac{1}{3}~~~~~~~\Longrightarrow~~~~~~~~A(J,\d)<\frac{1}{2 e 3^{3/2}}
\ee
Rough estimates give that (\ref{akj}) is satisfied when
\be{radius}
e^{2J}>4e3^{3/2}, \quad |\d|<\frac{1}{12e^{ 4J}}
\ee
this proves part $i)$ of Lemma \ref{main} in the $+$ b.c. case.
%%%%%%

%%%%
\noindent
The proof for periodic boundary conditions is { nearly} identical. { The only difference
is that now the maps $\s\mapsto \G$ is two-to-one (i.e. $\s$ and $-\s$ yield the same $\G$).
Each $\G$ is, as before,
the union of p-connected contours $\g$ where the connection is the same as
in the + b.c. case, and denote by   ${\cal P}^{per}_\L$ the set of all contour, which now also includes   contours winding around the torus $\L^{per}$. Therefore, we can write
}

\be{effekappa}
\p^{per}_G((f^{per})^k)=\frac{2}{Z^{per}_G}e^{J|B_{\L}|}(1+\d e^{-4J})^{k|\L|}\Xi^{per(k)}_\L(J,\d)
\ee
where the factor $2$ is due to the fact that a  contour can be filled in two ways and
$$
\Xi^{per(k)}_\L(J,\d)=\sum_{n\ge 0}{1\over n!}\sum_{(\g_1,...,\g_n)\in ({\cal P}^{per}_\L)^n\atop \g_i\sim \g_j}\prod_{i=1}^{n}
\left[e^{-2J|\g_i|}\xi_k(\g_i)\right]
$$
{ This implies that
$$
{1\over |\L|}\ln \p^{per}_G((f^{per})^k)=-{\ln(Z^{per}_G/2)\over |\L|} + J{|B_{\L}|\over |\L|}+k\ln(1+\d e^{-4J})+ {1\over |\L|}\ln \Xi^{per(k)}_\L(J,\d)
$$
with ${1\over |\L|}\ln \Xi^{per(k)}_\L(J,\d)$ being absolutely convergent if the same condition  (\ref{fp2})  is satisfied. Therefore ${1\over |\L|}\ln \p^{per}_G((f^{per})^k)$
is analytic in $\d$ uniformly in $\L$ whenever $J$ and $\d$ satisfy, as before, the condition given in (\ref{radius}).}
\bigskip

\noindent
{\it Part ii)}

\noindent
The first order in $\d$ of $\p^*_G((f^*)^k),\; k=1,2,$ can be directly computed. {  Recalling
(\ref{piG}), we have}
$$\p^*_G((f^*)^k)=
{\sum_{\s\in \cX} w^*_G(\s)(\prod_{i\in \L}(1+\delta\phi^*_i))^k\over \sum_{\s\in \cX} w^*_G(\s)}
=
\frac{\sum_{\s\in\cX}w^*_G(\s)(1+k\sum_{i\in\L}\d\phi^*_i+O(\d^2))}{\sum_{\s\in\cX}w^*_G(\s)}=
$$
$$
=1+k\d\frac{\sum_{\s\in\cX}w^*_G(\s)\sum_{i\in\L}\phi^{ *}_i}{\sum_{\s\in\cX}w^*_G(\s)}+O(\d^2)= 1+k\d\sum_{i\in\L}\p_G^*(\phi^{ *}_i)+O(\d^2)
$$
{ Therefore, for $k=1,2$, we get}
$$
{1\over |\L|}\p^*_G((f^*)^k)= k\d {\sum_{i\in \L}\pi_G^*(\phi^*_i)\over |\L|}+O(\d^2)
$$
from which part $ii)$ easily follows.

\noindent
The proof of theorem \ref{t2} is similar, and even easier.
Here we list the few changes needed in this case.
We define
\be{irr1}
h^I_i(\s)=J(\s_{i^\downarrow}+\s_{i^\leftarrow}),
\ee
\be{irr2}
{ \phi}^I_i=e^{-2(h^I_i(\s)\s_i}
\ee
and

\be{firr}
f^I(\sigma)=\prod_{i\in \L}(1+\delta\phi^I_i)
\ee
The theorem follows exactly in the same way if we prove
\bl{mainirr}
There exists $J_c$ such that for  any $J>J_c$
\bi
\item[i)] \hskip 1cm
$\displaystyle\frac{\ln \p^{per}_G((f^I)^2)}{|\L |}$ and $\displaystyle\frac{\ln \p^{per}_G(f^I)}{|\L |}$ are analytical functions of $\d$
for $|\d|<\d_J$.
\item[ii)] \hskip 1cm
$\displaystyle\frac{\ln \p^{per}_G((f^I)^2)}{|\L |}-2\displaystyle\frac{\ln \p^{per}_G(f^I)}{|\L |}=O(\d^2)$
\ei
\el
The proof  of the lemma uses the same ideas. We have

\be{effekappairr}
\p^{per}_G((f^I)^k)=\frac{1}{Z^{per}_G}e^{J|B_{\L}|}(1+\d e^{-4J})^{k|\L|}\sum_{\G\in {\cal G}_\L}
\left[e^{-2J|\G|}\xi^I_k(\G)\right]
\ee
where { now, due to (\ref{irr1}), (\ref{irr2}) }
\be{xiirr}
\xi^I_k(\G)=
\left[\frac{1+\d }{1+\d e^{-4J}}\right]^{k|l_1(\G)|}
\left[\frac{1+\d e^{+4J}}{1+\d e^{-4J}}\right]^{k|l_2(\G)|}
\ee
Then we split $\G$ in its connected components $\g_1,...,\g_n$, where the notion of connection
is in this case the usual one { (since if $i\in l_2(\G)$ then, due to  (\ref{irr1}), the two segments of $\G$ at distance $1/2$ from $i$ must form an elbow.).
Denoting by ${\cal P}^I_\L$  (by $ {\cal P}^I$) the set of standard contours in $\L$  (in $\mathbb{Z}^d$)}, we can therefore write
$$
\sum_{\G\in {\cal G}_\L}
e^{-2J|\G|}\xi^I_k(\G)= \Xi^{I(k)}_\L(J,\d)=\sum_{n\ge 0}{1\over n!}\sum_{(\g_1,...,\g_n)\in ({\cal P}^I_\L)^n\atop \g_i\sim \g_j}\prod_{i=1}^{n}
\left[e^{-2J|\g_i|}\xi^I_k(\g_i)\right]
$$
{ So, as before, we have to analyze the absolute convergence of the logarithm of the quantity $\Xi^{I(k)}_\L(J,\d)$, the grand-canonical partition function
of a a hard core polymer gas in which polymers are now standard contours in the set ${\cal P}_\L$ with
activity}
$$
\r^I_k(\g)=\xi^I_k(\g)e^{-2J|\g|}
$$
and we have to prove that
\be{fpirr}
\sum_{\g\in {\cal P}'\atop x
\in \g}|\r^I_k(\g)|e^{|\g|} \le 1
\ee
{ where $\cal  P$ is now the set of usual contours in $\mathbb{Z}^2$ and  $x$ is  any fixed vertex of $\mathbb{Z}^2$}, due to the translation invariance of the model.
We can easily find a bound for $|\r^I_k(\g)|$: we get,
for $|\d|<e^{4J}$
\be{st1irr}
\r^I_k(\g)\le e^{-2J|\g|}\left[\frac{1+|\d| e^{+4J}}{1-|\d| e^{-4J}}\right]^{k\sum_{s=1}^2|l_s(\g)|}
\ee
Observing that in this case
$$
\sum_{s=1}^2|l_s(\g)|\le |\g|
$$
we get, { for any $k=1,2$}
\be{st2irr}
\r^I_k(\g)\le e^{-2J|\g|}\left[\frac{1+|\d| e^{+4J}}{1-|\d| e^{-4J}}\right]^{k|\g|}  \le e^{-2J|\g|}\left[\frac{1+|\d| e^{+4J}}{1-|\d| e^{-4J}}\right]^{2|\g|}
\ee
We call then
$$
 A(J,\d)=e^{-2J}\left[\frac{1+|\d| e^{+4J}}{1-|\d| e^{-4J}}\right]^{2}
$$
and we obtain
$$
\sum_{\g\in {\cal P}'\atop x
\in \g}\r^I_k(\g)e^{|\g|}\le \sum_{n\ge 4}A^I(J,\d)^n e^n
\sum_{\g\ni x\atop |\g|=n}1
\le
\sum_{n\ge 4}[3eA^I(J,\d)]^n
$$
Here the factor $3^n$ comes form the usual estimate on Peierls contour.
The rest of the proof is identical.
\vskip1.truecm
\noindent
{\bf Acknowledgments:}
It is a pleasure to thank Giosi Benfatto for useful discussions. A.P.  has been partially supported by the Brazilian  agencies
Conselho Nacional de Desenvolvimento Cient\'\i fico e Tecnol\'ogico
(CNPq) and  Funda\c c\~ao de Amparo \`a  Pesquisa do Estado de Minas Gerais (FAPEMIG - Programa de Pesquisador Mineiro).
B.S. and E.S. thank
 the support of the A*MIDEX project (n. ANR-11-IDEX-0001-02) funded by the  ``Investissements d'Avenir" French Government program, managed by the French National Research Agency (ANR).
B.S. has been supported by PRIN 2012, Problemi matematici in teoria cinetica ed
applicazioni.

%\vglue2.truecm
%{\bf Ho messo a posto  le referenze  in ordine alfabetica ordine alfabetica (e completato quelle incomplete).  I titoli dei 2 vostri lavori [3] e [4] erano invertiti!
%Ho aggiunto le ref, [1] [15] e [20]. }


\begin{thebibliography}{99}

{
\bibitem{BFP} {\sc R. Bissacot, R. Fern\'andez, A. Procacci}, {\em On the convergence of cluster expansions for polymer gases},
\newblock J.  Statist. Phys. {\bf 139} (4), 598-617 (2010).}

\bibitem{CN}
{\sc E.N.M.Cirillo, F.R.Nardi}
\newblock{\em Metastability for Stochastic Dynamics with a Parallel Heat Bath Updating Rule},
\newblock J.  Statist. Phys. {\bf 110}, 183-217 (2003).

\bibitem{CNS}
{\sc E.N.M.Cirillo, F.R.Nardi, C.Spitoni}
\newblock{\em Metastability ffor reversible Probabilistic Cellular Automata with self-interacton},
\newblock J.  Statist. Phys. {\bf 132}, 431-471 (2008).

\bibitem{dss1}
{\sc P.Dai Pra, B.Scoppola, E.Scoppola}
\newblock{\em  Sampling from a Gibbs measure with pair interaction by means of PCA}
\newblock J.  Statist. Phys. {\bf 149}, 722-737 (2012).

\bibitem{dss2}
{\sc P.Dai Pra, B.Scoppola, E.Scoppola}
\newblock{\em  Fast mixing for the low temperature 2D Ising model through irreversible parallel dynamics}
\newblock J.  Statist. Phys. {\bf 159}, 1-20 (2015).


\bibitem{D}
{\sc D. Dawson}, \newblock{\em A.,Synchronous and asynchronous reversible Markov systems}, \newblock Canad. Math. Bull. {\bf 17}
 no. 5, 633-649 (1974/75).

 \bibitem{fp}
{\sc R.Fernandez, A. Procacci}
\newblock{\em Cluster expansion for abstract polymer models.New bounds from an old approach}
\newblock Commun. Math. Phys. {\bf 274}, 123-140 (2007).

\bibitem{FT}
{\sc R. Fernandez, A. Toom}, \newblock{\em Non Gibbsiannes of the invariant measures of non-reversible cellular automata with totally asymmetric noise}, \newblock Ast\'erisque {\bf 287}, 71-87 (2003).

\bibitem{GSSV}
{\sc A.Gaudilli\`ere, B.Scoppola, E.Scoppola, M.Viale }, \newblock{\em Phase transition for the cavity approach to the clique problem on random graphs.},
\newblock J.  Statist. Phys.  { {\bf 145}, n. 5, 1127-1155 (2011)}.


\bibitem{GKLM}
{\sc S. Goldstein , R. Kuik , J.L. Lebowitz, C. Maes}, \newblock{\em From PCAs to equilibrium systems and back.},
\newblock Commun. Math. Phys. {\bf 125}, no. 1, 71-79 (1989).

\bibitem{IN}
{\sc J. Z. Imbrie1, C. M. Newman}, \newblock {\em An Intermediate Phase with Slow Decay of Correlations in One Dimensional $1/|i-j|^2$
Percolation, Ising and Potts Models}, \newblock Commun. Math. Phys. {\bf 118}, 303--336 (1988).

\bibitem{ISS}
{\sc A.Iovanella, B.Scoppola, E.Scoppola}
{
\newblock{\em Some Spin Glass Ideas Applied to the Clique Problem},
\newblock J.  Statist. Phys.  {\bf 126}, n. 4, 895-915 (2007)}.

\bibitem{kp}
{\sc R. Koteck\'y, D.Preiss}, \newblock{\em Cluster expansion for abstract polymer models}, \newblock Commun. Math. Phys. {\bf 103}, 491--498 (1986).


\bibitem{KV}
{\sc O. Kozlov, N. Vasilyev}, \newblock{\em Reversible Markov chains with local interaction, Multicomponent random
systems}, \newblock Adv. Probab. Related Topics {\bf 6}, Dekker, New York, pp. 451-469 (1980).

\bibitem{Ku82}
{\sc  H. K\"{u}nsch}, {\em Decay of Correlations under Dobrushin's Uniqueness
Condition and its Applications}, Commun. Math. Phys. {\bf 84}, 207-222 (1982).

{ \bibitem{LS}
{\sc C. Lancia, B. Scoppola}, {\em Equilibrium and Non-equilibrium Ising Models
by Means of PCA}
\newblock
 J. Statist. Phys. {\bf 153},  641-653  (2013)}.

\bibitem{LMS}
{\sc J.L. Lebowitz, C. Maes, E.R. Speer}, \newblock{\em Statistical mechanics of probabilistic cellular automata}, \newblock J.
Statist. Phys. {\bf 59}, no. 1-2, 117-170  (1990).

\bibitem{LLP}
{\sc D.A.Levin, M.J.Luczak, Y.Peres}
\newblock{\em Glauber dynamics for the mean-field Ising model: cut-off, critical power law and metastability},
{ \newblock Probab. Theory Related Fields {\bf 146}, 223-265 (2010)}.

 \bibitem{M}
{\sc F.Martinelli}
\newblock{\em Relaxation time of Markoc chains in statistical mechanics and combinatorial structures}
\newblock in ``Probability on discrete structures", Encyclopedia of Mathematical Sciences, vol. 110 - Springer  (2004).

\bibitem{N}
{\sc C.M. Newman}, \newblock{\em A general central limit theorem for FKG systems}, \newblock Commun. Math. Phys. {\bf 91}, n. 1, 75-80,  (1983).

\bibitem{OV}{ {\sc   E. Olivieri, M. E. Vares}, \newblock{\em Large deviations and metastability}. \newblock Encyclopedia of Mathematics and its Applications,
100. Cambridge University Press, Cambridge, (2005)}.

\bibitem{SW}
{\sc J.S.Wang, R.H.Swendsen}
\newblock{\em  Non universal critical dynamics in Monte Carlo simulations}
\newblock Phys. Rev. Lett. {\bf 58}, 86-88 (1987).


\end{thebibliography}
\end{document}